\documentstyle[12pt,dina4p]{article}
\newcommand{\be}{\begin{equation}}
\newcommand{\ee}{\end{equation}}
\newcommand{\ba}{\begin{eqnarray}}
\newcommand{\ea}{\end{eqnarray}}
\newcommand{\bc}{\begin{center}}
\newcommand{\ec}{\end{center}}

\newcommand{\bay}{\begin{array}{rcl}}
\newcommand{\eay}{\end{array}}
\newcommand{\dis}{\displaystyle}
\newcommand{\text}{\textstyle}

\newcommand{\rf}[1]{(\ref{#1})}

\renewcommand{\arraystretch}{2.0}
  \newcommand{\mysection}[1]{\section{#1}\setcounter{figure}{0}
                    \setcounter{table}{0}\setcounter{equation}{0}}
\def\Tr{\mbox{\rm Tr}}
\def\tr{\mbox{\rm tr}}

\def\DD{D\kern-0.82em\lower-.2ex\hbox{$\not$}\kern+0.8em}

\def\gb{\bar{g}}
\def\hb{\bar{h}}
\def\rb{\bar{R}}
\def\gb{\bar{g}}

\def\wh{\widehat}

\def\fs{
   \mbox{$ F_{\mu\nu}^a \kern -1.65em  \lower -.2ex \hbox{${}^*$}$}
     \kern +1.5em
       }
\def\fsn{
   \mbox{$ F \kern -0.9em  \lower -.2ex \hbox{${}^*$}$}
     \kern +0.35em
       }

\def\ca{\mbox{${\cal A}$}}
\def\cd{\mbox{${\cal D}$}}
\def\cf{\mbox{${\cal F}$}}
\def\cj{\mbox{${\cal J}$}}
\def\cl{\mbox{${\cal L}$}}
\def\cm{\mbox{${\cal M}$}}
\def\cn{\mbox{${\cal N}$}}
\def\co{\mbox{${\cal O}$}}
\def\cs{\mbox{${\cal S}$}}
\def\cz{\mbox{${\cal Z}$}}
\def\gb{\bar{g}}
\def\mn{{\mu\nu}}
\def\gmn{\gamma_\mn}
\def\znk{\mbox{$Z_{Nk}$}}
\begin{document}
\noindent
\begin{titlepage}
\begin{flushright}
       DESY 96--080\\
       hep-th/9605030
\end{flushright}
  \vspace{1.0in} \LARGE
  \begin{center}
    {\bf Nonperturbative Evolution Equation for \\
      Quantum Gravity
     }\\ \vspace{0.3in}
  \end{center}
  \vspace{0.2in} \large
  \begin{center}
    {\bf {M. Reuter} \\
     \it {Deutsches Elektronen-Synchrotron DESY\\
     Notkestrasse 85  \\
     D-22603 Hamburg   \\
     Germany}
     }\\ \vspace{0.3in}
  \end{center}
  \vspace{1.0in}
 \normalsize

\begin{abstract}
A scale--dependent effective action for gravity is introduced and
an exact nonperturbative evolution equation is derived which governs
its renormalization group flow. It is invariant under general coordinate
transformations and satisfies modified BRS Ward--Identities. The
evolution equation is solved for a simple truncation of the space of
actions. In $2+\varepsilon$ dimensions, nonperturbative corrections to
the $\beta$--function of Newton's constant are derived and its dependence on
the cosmological constant is investigated. In 4 dimensions, Einstein
gravity is found to be ``antiscreening'', i.e., Newton's constant
increases at large distances.
\end{abstract}

\end{titlepage}

\mysection{Introduction}
In many of the traditional approaches to quantum gravity the
Einstein--Hilbert term has been regarded as a fundamental action which
should be quantized along the same lines as the familiar renormalizable
field theories in flat space, such as QED for example \cite{pqg}.
It was soon realized that this program is not only technically rather
involved but also leads to severe conceptual difficulties. In particular,
the nonrenormalizability of the theory hampers a meaningful perturbative
analysis.
While this does not rule out the possibility that the theory exists
nonperturbatively, not much is known in this direction. However, it could
also be argued that gravity, as we know it, should not be quantized at
all, because Einstein gravity is an effective theory \cite{effrev}
which results from quantizing some yet unknown fundamental theory.
If so, the Einstein--Hilbert term is an effective action analogous to
the Heisenberg--Euler action in QED and it should not be compared to
the ``microscopic'' action of electrodynamics.

It seems not unreasonable to assume that the truth lies somewhere
between those two extreme points of view, i.e., that Einstein gravity
is an effective theory which is valid near a certain nonzero momentum
scale $k$. This means that it arises from the fundamental theory by a
``partial quantization'' in which only excitations with momenta
larger than $k$ are integrated out, while those with momenta smaller
than $k$ are not included.
(The interpretation of the Einstein--Hilbert term as a fundamental or
an ordinary effective action is recovered in the limits $k\to \infty$
and $k\to 0$, respectively.)
An ``effective theory at scale $k$'', when evaluated at tree level,
should correctly describe all gravitational phenomena which involve
a typical momentum scale $k$ acting as a physical infrared cutoff.
Only if one is interested in processes with momenta $k^\prime\ll k$,
loop calculations become necessary; they amount to integrating out
the missing field modes in the momentum interval $[k^\prime,k]$.

We shall regard the scale--dependent action for gravity, henceforth
denoted $\Gamma_k$, as a Wilsonian effective action which is obtained
from the fundamental (``microscopic'') action $S$ by a kind of
coarse--graining analogous to the iterated block--spin transformations
which are familiar from lattice systems \cite{wil}. In the
continuum, $\Gamma_k$ will be defined in terms of a modified
functional integral over $e^{-S}$ in which the contributions of all
field modes with momenta smaller than $k$ are suppressed. In this
manner $\Gamma_k$ interpolates between $S$ (for $k\to \infty$) and
the effective action $\Gamma$ (for $k\to 0$).
The trajectory in the space of all action functionals can be obtained
as the solution of a certain functional evolution equation, the exact
renormalization group equation. Its form is independent of the action
$S$ under consideration.
The latter enters via the initial conditions for the renormalization
group trajectory; it is specified at some UV cutoff scale $\Lambda$:
$\Gamma_\Lambda = S$. If $S$ is a truly fundamental action, $\Lambda$
is sent to infinity at the end. The renormalization group equation
can also be used to evolve effective actions, known at some point
$\Lambda$, towards smaller scales $k<\Lambda$. In this case $\Lambda$
is a fixed, finite scale. In this framework, the (non)renormalizability
of a theory is seen as a global property of the renormalization
group flow for $\Lambda\to\infty$. The evolution equation by itself
is perfectly finite and well behaved in either case, because it
describes only infinitesimal changes of the cutoff.

In this paper we shall give a precise meaning to the notion of a
scale--dependent gravitational action $\Gamma_k[g_\mn]$ and we shall
derive the associated evolution equation. We employ a formulation
in which the metric is the fundamental dynamical variable. Alternative
approaches based upon the spin--connection and the vielbeins are
also possible, but they will not be considered here.
By using a variant of the background gauge technique we are able
to make $\Gamma_k[g_\mn]$ invariant under general coordinate
transformations. This property is very important if one wants to find
nonperturbative solutions of the evolution equations in terms of simple
truncations of the space of actions.
Our construction of $\Gamma_k [g_\mn]$ parallels the definition
of the ``effective average action'' \cite{wet,rw3} which was widely
used recently \cite{ah,ng,cs,topren}.\footnote{
For related work using similar techniques see refs.
\cite{exrg,mor,ehw}.}
The remarkable successes of this method in flat space are partly due
to the fact that it allows for nonperturbative solutions when no
small expansion parameter is available, and that
$\Gamma_k$ has a built--in infrared cutoff. Therefore the low--momentum
behavior of (almost) massless theories can be investigated even in
cases where IR divergences render standard perturbation theory
inapplicable. For the purposes of quantum gravity, both of these
features are very welcome, of course. In fact, in quantum cosmology
one of the most intriguing questions is how quantized Einstein
gravity behaves at extremely large distances. It has been argued
\cite{tsam,mott} that in presence of a nonzero cosmological constant
there should be very strong renormalization effects in the infrared
which might even provide a mechanism for a dynamical relaxation
of the cosmological constant. The method which we are going to
develop would be ideally suited to study problems of this type.
Since only long distance physics is involved here, there are good
chances that this can be done without knowing the microscopic
theory of quantum gravity. (See ref. \cite{effrev} for a related
discussion.)

The ``effective average action'' used in this paper should not be
confused with the closely related ``average action'' which was introduced
earlier \cite{ring}. The former obeys a more convenient evolution
equation while the latter has a simple interpretation in terms of
field averages. Their precise relation is explained in ref.\cite{relat}.
The average action has been used in a gravitational context in
refs.\cite{bon}, \cite{perc}, but no exact evolution equation was
formulated. The evolution of the effective average action in a
gravitational background was studied in ref.\cite{liou} in the
context of Liouville field theory. For a review of the effective
average action and its applicaton to Yang--Mills theory we refer
to \cite{corfu}.

The remaining sections of this paper are organized as follows.
In section 2 we give the definition of $\Gamma_k$ and derive the
exact, nonperturbative renormalization group equation. In section 3
we establish the modified Ward identities satisfied by $\Gamma_k$,
and we show that the conventional diffeomorphism Ward identities
are recovered in the limit $k\to 0$.
In its general form, the evolution equation describes a flow on the
infinite dimensional space of all action functionals. Approximate
nonperturbative solutions can be found by truncating the space of
actions, i.e., by projecting the flow on a finite--dimensional
subspace. In section 4 we investigate the ``Einstein--Hilbert truncation''
where only the operators $\int \sqrt{g}$ and $\int \sqrt{g} R$
are retained. In section 5 we determine the resulting
scale dependence of
Newton's constant and of the cosmological constant. As an example,
gravity in $2+\varepsilon$ and in 4 dimensions is discussed in detail.

\mysection{The Renormalization Group Equation}
In this section we introduce the effective average action for
euclidean quantum gravity in $d$ dimensions
and
we derive the exact renormalization group equation which governs
its scale dependence.

We are going to employ the background gauge fixing technique
\cite{adl,odin} which means that we decompose the integration
variable $\gamma_{\mu\nu}(x)$ in the functional
integral over all metrics according to
\be\label{2.1}
\gamma_{\mu\nu}(x)=\bar{g}_{\mu\nu}(x)+h_{\mu\nu}(x)
\ee
Here $\bar{g}_{\mu\nu}$ is a fixed background metric so that the
integration over $\gamma_{\mu\nu}$ may be replaced by an integration
over $h_{\mu\nu}$. We consider the following scale--dependent
modification of the generating functional for the connected
Green's functions
\be\label{2.2}
\begin{array}{c}
\dis
\exp\left\{
      W_k
       \left[t^{\mu\nu},\sigma^\mu,\bar{\sigma}_\mu;
             \beta^{\mu\nu},\tau_\mu;\bar{g}_{\mu\nu}
       \right]
    \right\}
\qquad
\qquad
\qquad
\qquad
\qquad
\qquad
\qquad
\\
\hfill
\dis
\qquad
\qquad
\qquad
=\int \cd h_{\mu\nu}\cd C^\mu \cd\bar{C}_\mu
\,
\exp\Big\{
     -S[\bar{g}+h]-S_{\rm gf}[h;\bar{g}]
\\
\qquad
\qquad
\qquad
\qquad
\qquad
\qquad
\dis
     -S_{\rm gh}[h,C\bar{C};\bar{g}]-\Delta_kS[h,C,\bar{C};\bar{g}]
     -S_{\rm source}
      \Big\}
\end{array}
\ee
Here $S[\gamma]=S[\gb+h]$ is the classical action which
is assumed to be invariant under the general coordinate
transformations
\be\label{2.3}
\delta\gmn=\cl_{v}\gmn \equiv v^\rho\partial_\rho \gmn
+\partial_\mu v^\rho \gamma_{\rho\nu}
+\partial_\nu v^\rho \gamma_{\mu\rho}
\ee
where $\cl_{v}$ denotes the Lie derivative with respect to the
vector field $v^\mu$. For the time being let us also assume that
$S$ is positive definite.

Furthermore, $S_{\rm gf}$ denotes the gauge fixing term for the
gauge condition $F_\mu(\gb,h)=0$,
\be\label{2.4}
S_{\rm gf}[h;\gb]=\frac{1}{2\alpha}\int \, d^dx
\sqrt{\gb}\,\gb^\mn F_\mu F_\nu
\ee
and $S_{\rm gh}$ is the action for the corresponding Faddeev--Popov
ghosts $C^\mu$ and $\bar{C}_\mu$:
\be\label{2.5}
S_{\rm gh}[h,C,\bar{C};\gb]=
-\kappa^{-1}\int \, d^dx \,\bar{C}_\mu\, \gb^\mn
\,
\frac{\partial F_\nu}{\partial h_{\alpha\beta}}
\,\cl_C\left(\gb_{\alpha\beta}+h_{\alpha\beta}\right)
\ee
The Faddeev--Popov action $S_{\rm gh}$ is obtained along the
same lines as in Yang--Mills theory: one applies a gauge
transformation
\renewcommand{\arraystretch}{1.3}
\be\label{2.6}
\begin{array}{rcl}
\dis
\delta h_\mn&=&\dis\cl_v\gmn = \cl_v(\gb_\mn+h_\mn)
\\
\dis
\delta \gb_\mn & =&\dis 0
\eay
\ee
to $F_\mu$ and replaces the parameters $v^\mu$ by the ghost
field $C^\mu$. The integral over $C^\mu$ and $\bar{C}_\mu$ provides
a representation of the Faddeev--Popov determinant
$\det[\delta F_\mu/\delta v^\nu]$ then. In eq. \rf{2.5} we introduced
the constant (proportional to the Planck mass)
\be\label{2.7}
\kappa\equiv\left(32\pi\bar{G}\right)^{-1/2}
\ee
where $\bar{G}$ denotes the bare Newtonian constant.
In principle our construction works for an arbitrary
background gauge fixing. It is particularly convenient
to use a $F_\mu$ which is linear in the quantum field $h_\mn$:
\be\label{2.8}
F_\mu=\sqrt{2}\kappa\, \cf_\mu^{\alpha\beta}\left[\gb\right]
h_{\alpha\beta}
\ee
We shall mostly employ the harmonic coordinate condition
for which $\cf_\mu^{\alpha\beta}$ is the following first order
differential operator constructed from $\gb_\mn$:
\be\label{2.9}
\cf_\mu^{\alpha\beta}=\delta_\mu^\beta \gb^{\alpha\gamma}
\bar{D}_\gamma-\frac{1}{2}\gb^{\alpha\beta}\bar{D}_\mu
\ee
The covariant derivative $\bar{D}_\mu$ involves the Christoffel
symbols $\bar{\Gamma}_\mn^\rho$ of the background metric $\gb_\mn$.
For the gauge fixing \rf{2.8} with \rf{2.9} the ghost action
reads
\be\label{2.9a}
S_{\rm gh}[h,C,\bar{C};\gb]=-\sqrt{2}\int d^dx \,  \sqrt{\gb}
\,\bar{C}_\mu \cm [g,\gb]^\mu{}_\nu C^\nu
\ee
with the Faddeev--Popov operator
\be\label{2.9b}
\cm[g,\gb]^\mu{}_\nu=
\gb^{\mu\rho} \gb^{\sigma\lambda}
\bar{D}_\lambda(g_{\rho\nu} D_\sigma +g_{\sigma\nu}D_\rho)
-\gb^{\rho\sigma}\gb^{\mu\lambda}\bar{D}_\lambda g_{\sigma\nu} D_\rho
\ee

The essential piece in eq.\rf{2.2} is the IR cutoff for the gravitational
field $h_\mn$ and for the ghosts:
\renewcommand{\arraystretch}{1.6}
\be\label{2.10}
\bay
\dis
\Delta_k S[h,C,\bar{C};\gb]
&=&\dis
\frac{1}{2}\kappa^2\int d^dx \, \sqrt{\gb}\, h_\mn
R^{\rm grav}_k[\gb]^{\mn\rho\sigma}h_{\rho\sigma}
\\
&&\dis
 +\sqrt{2}\int d^dx\, \sqrt{\gb}\, \bar{C}_\mu R^{\rm gh}_k[\gb]C^\mu
\eay
\ee
The cutoff operators $R^{\rm grav}_k$ and $R^{\rm gh}_k$ serve the purpose
of discriminating between high--momentum and low--momentum modes.
Eigenmodes of $-\bar{D}^2$ with eigenvalues $p^2\gg k^2$ are integrated out
in \rf{2.2} without any suppression whereas modes with small eigenvalues
$p^2\ll k^2$ are suppressed by a kind of momentum dependent mass term.
The operators $R^{\rm grav}_k$ and $R^{\rm gh}_k$ describe the transition
from the high--momentum to the low--momentum regime. Either of them
has the structure
\be\label{2.11}
R_k[\gb]=\cz_k k^2 R^{(0)}(-\bar{D}^2/k^2)
\ee
where the dimensionless function $R^{(0)}$ interpolates smoothly
between $R^{(0)}(0)=1$ and $\dis \lim_{u\to \infty} R^{(0)}(u)=0$.
A convenient choice is for example
\be\label{2.11a}
R^{(0)}(u)=u[\exp(u)-1]^{-1}
\ee
The factors $\cz_k$ are different for the graviton and the ghost cutoff.
For the ghost $\cz_k\equiv Z^{\rm gh}_k$ is a pure number, whereas for
the metric fluctuation $\cz_k\equiv \cz^{\rm grav}_k$ is a
tensor constructed from the background metric $\gb_\mn$.
In the simplest case one would take
\be\label{2.10a}
\left(\cz_k^{\rm grav}\right)^{\mn \rho \sigma}
=\gb^{\mu \rho} \gb^{\nu \sigma} Z^{\rm grav}_k
\ee
In section 4 we shall employ a slightly more refined choice. There we
shall also explain how the factors $Z^{\rm gh}_k$ and $Z^{\rm grav}_k$
should be choosen.
Note that the cutoff action \rf{2.10} is quadratic in the
quantum fields $h_\mn$, $C^\mu$ and $\bar{C}_\mu$. This is an important
prerequisite for obtaining a tractable evolution equation later on.
The requirement of a quadratic $\Delta_k S$ forces
us to use the covariant Laplacian
$\bar{D}^2\equiv \gb^\mn \bar{D}_\mu\bar{D}_\nu$
in the {\it background} metric
as the operator which discriminates between
high--momentum and low--momentum modes.

In \rf{2.2} we coupled $h_\mn$, $C^\mu$ and $\bar{C}_\mu$ to the sources
$t^\mn$, $\bar{\sigma}_\mu$ and $\sigma^\mu$, respectively:
\be\label{2.12}
\bay\dis
S_{\rm source}
& =&\dis
 -\int d^dx \, \sqrt{\gb}
\Big\{ t^\mn h_\mn +\bar{\sigma}_\mu C^\mu +\sigma^\mu \bar{C}_\mu
\\
&& \dis \qquad \qquad \quad
+\beta^\mn \cl_C \left( \gb_\mn +h_\mn\right)
               +\tau_\mu C^\nu \partial_\nu C^\mu
\Big\}
\eay
\ee
The sources $\beta^\mn$ and $\tau_\mu$ couple to the BRS variations
of $h_\mn$ and $C^\mu$, respectively.
In fact, it is not difficult
verify that $S+S_{\rm gf}+S_{\rm gh}$ is invariant under the
BRS transformations ($\varepsilon$ is an anticommuting parameter)
\renewcommand{\arraystretch}{1.4}
\be\label{2.13}
\bay
\dis
\delta_\varepsilon
h_\mn
&=& \dis
\varepsilon \kappa^{-2} \cl_C \gamma_\mn
=\varepsilon \kappa^{-2} \cl_C\left(\gb_\mn+h_\mn\right)
\\
\dis
\delta_\varepsilon
\gb_\mn &=& \dis 0
\\
\delta_\varepsilon
C^\mu
&=& \dis
\varepsilon \kappa^{-2} C^\nu \partial_\nu C^\mu
\\
\delta_\varepsilon
\bar{C}_\mu
&=& \dis
\varepsilon \left(\alpha\kappa\right)^{-1} F_\mu
\eay
\ee

Given the functional $W_k$, we introduce $k$--dependent
classical fields
\be\label{2.14}
\bar{h}_\mn = \frac{1}{\sqrt{\gb}}\frac{\delta W_k}{\delta t^\mn}
\qquad , \qquad
\xi^\mu=\frac{1}{\sqrt{\gb}}\frac{\delta W_k}{\delta \bar{\sigma}_\mu}
\qquad , \qquad
\bar{\xi}_\mu=\frac{1}{\sqrt{\gb}}\frac{\delta W_k}{\delta\sigma^\mu}
\ee
and we formally solve for the sources
$(t^\mn\, , \, \sigma^\mu \, , \, \bar{\sigma}_\mu )$ as functionals
of the fields
$(\bar{h}_\mn \, , \, \xi^\mu \, , \, \bar{\xi}_\mu )$ and of
$(\beta^\mn \, , \, \tau_\mu \, ; \, \bar{g}_\mn)$.
Then the Legendre transform $\widetilde{\Gamma}_k$ of $W_k$ depends on the
classical fields and parametrically on $\beta$, $\tau$ and $\gb$:
\be\label{2.15}
\widetilde{\Gamma}_k[\bar{h},\xi,\bar{\xi}; \beta, \tau ; \gb]
=
\int d^dx \, \sqrt{\gb}
\left\{
t^\mn \bar{h}_\mn +\bar{\sigma}_\mu \xi^\mu + \sigma^\mu\bar{\xi}_\mu
\right\}
-W_k[t,\sigma,\bar{\sigma}; \beta,\tau ; \gb]
\ee
By definition, the effective average action $\Gamma_k$ obtains from
$\widetilde{\Gamma}_k$ by subtracting the cutoff action $\Delta_k S$ with the
classical fields inserted:
\be\label{2.16}
\Gamma_k[\bar{h},\xi,\bar{\xi}; \beta, \tau ; \gb]
=
\widetilde{\Gamma}_k[\bar{h},\xi,\bar{\xi}; \beta, \tau ; \gb]
-
\Delta_k S[\bar{h}, \xi , \bar{\xi} ; \gb]
\ee
It is convenient to define the metric
\be\label{2.16a}
g_\mn(x)\equiv \gb_\mn (x) +\bar{h}_\mn  (x)
\ee
as the classical analogue of the quantum metric
$\gamma_\mn \equiv \gb_\mn +h_\mn$ and to consider $\Gamma_k$ as a
functional of $g_\mn$ rather than $\bar{h}_\mn$:
\be\label{2.17}
\Gamma_k[g_\mn ,\gb_\mn , \xi^\mu , \bar{\xi}_\mu ; \beta , \tau ]
\equiv
\Gamma_k[g_\mn -\gb_\mn , \xi^\mu , \bar{\xi}_\mu ; \beta , \tau ; \gb_\mn]
\ee
The main virtue of the background technique employed here is that the
functional $\Gamma_k$ is invariant under general coordinate transformations
where all its arguments transform as tensors of the corresponding rank:
\be\label{2.18}
\Gamma_k[\Phi +\cl_v \Phi] = \Gamma_k[\Phi]
\qquad , \qquad
\Phi\equiv
\left\{
  g_\mn , \gb_\mn , \xi^\mu , \bar{\xi}_\mu ; \beta^\mn , \tau_\mu
\right\}
\ee
Note that in \rf{2.18}, contrary to the ``gauge transformation''
\rf{2.6}, also the background metric transforms as an ordinary
tensor field:
$\delta\gb_\mn = \cl_v \gb_\mn$.
Eq. \rf{2.18} is a consequence of
\be\label{2.19}
W_k\left[\cj + \cl_v \cj\right] = W_k\left[\cj\right]
\quad , \quad
\cj \equiv
\left\{
  t^\mn , \sigma^\mu , \bar{\sigma}_\mu ;
  \beta^\mn, \tau_\mu; \gb_\mn
\right\}
\ee
This invarianc property follows from \rf{2.2} if one performs
a  compensating transformation on the integration variables
$h_\mn$, $C^\mu$ and $\bar{C}_\mu$. At this point we assume
that the measure is diffeomorphism invariant.

The general coordinate invariance of $\Gamma_k$ is of major
practical importance because if we know a priori that no
symmetry--violating terms are generated during the evolution
it is sufficient to use truncations which consist of invariant
combinations of the fields only. The conventionally defined effective
action of the metric, $\Gamma[g_\mn]$, obtains
in the limit of a vanishing IR cutoff by
setting the ghosts, $\beta$ and $\tau$ to zero and by identifying
$\gb_\mn$ with $g_\mn$:
\be\label{2.20}
\Gamma[g_\mn]
=\lim_{k \to 0}
\Gamma_k[g_\mn, g_\mn,0,0;0,0]
\ee
As a consequence, $\Gamma[g_\mn]$ is invariant under
$\delta g_\mn =\cl_v g_\mn$.
Even though we are mostly interested
in the functional
\be\label{2.21}
\bar{\Gamma}_k[g_\mn]\equiv
\Gamma_k[g_\mn,g_\mn,0,0;0,0]
\ee
which depends on $g_\mn$ only, an exact renormalization
group equation can be formulated only if one keeps
track of the dependence on $\xi$, $\bar\xi$ and $\gb$ as well.
For the derivation of the (modified) BRS Ward identities
satisfied by $\Gamma_k$ the dependence on $\beta$ and $\tau$ must
be retained in addition.

The derivation of the evolution equation for $\Gamma_k$ proceeds as
follows. Taking a derivative of the functional integral \rf{2.2}
with respect to the renormalization group ``time''
$t\equiv \ln k$ one obtains, in matrix notation
\be\label{2.22}
-\partial_t W_k
=
\frac{1}{2} \Tr
\left[ < h\otimes h> \left(\partial_t \widehat{R}_k\right)_{\bar{h}\bar{h}}
\right]
- \Tr
\left[ < \bar{C}\otimes C>
\left(\partial_t \widehat{R}_k\right)_{\bar{\xi}\xi}
\right]
\ee
Here $\widehat{R}_k$ is a matrix in field space whose non--zero entries are
\renewcommand{\arraystretch}{2.0}
\be\label{2.23}
\bay\dis
\left(\widehat{R}_k\right)^{\mn \rho\sigma}_{\bar h \bar h}
&=& \dis
\kappa^2
\left(R_k^{\rm grav}[\gb]\right)^{\mn\rho\sigma}
\\
\left(\widehat{R}_k\right)_{\bar \xi \xi}
&=& \dis
\sqrt{2}
R^{\rm gh}_k[\gb]
\eay
\ee
The RHS of \rf{2.22} can be expressed in terms of $\Gamma_k$
by noting that the connected two--point function
\be\label{2.24}
\bay\dis
G_{ij}(x,y)
&\equiv&\dis
<\chi_i(x) \chi_j(y)> -\varphi_i(x)\varphi_j(y)
\\
&=&\dis
\frac{1}{\sqrt{\gb(x)\gb(y)}}
\,
\frac{\delta^2 W_k}{\delta J^i(x)\delta J^j(y)}
\eay
\ee
and
\be\label{2.25}
\widetilde\Gamma^{(2)\, ij}_k(x,y)
\equiv
\frac{1}{\sqrt{\gb(x)\gb(y)}}
\,
\frac{\delta^2 \widetilde\Gamma_k}
     {\delta \varphi_i(x)\delta \varphi_j(y)}
\ee
are inverse matrices in the sense that
\be\label{2.26}
\int d^dy\,
\sqrt{\gb(y)}\,
G_{ij}(x,y)
\,
\widetilde\Gamma^{(2)\,jl}_k(y,z)
=\delta^l_i\,\frac{\delta(x-z)}{\sqrt{\gb(z)}}
\ee
Here we used the shorthand notation
$\chi_i\equiv \{ h, C ,\bar C \}$,
$J^i\equiv \{t,\sigma,\bar\sigma \}$
and
$\varphi_i\equiv \{\bar h,\xi,\bar\xi \}$.
Thus one obtains the evolution equation
\be\label{2.27}
\begin{array}{rcl}
\dis
\partial_t \Gamma_k[\bar h, \xi, \bar\xi; \beta, \tau; \gb] &=
&
\dis
\frac{1}{2}\Tr
\left[\left(\Gamma^{(2)}_k + \widehat R_k\right)^{-1}_{\bar h \bar h}
\left(\partial_t \widehat R_k\right)_{\bar h \bar h} \right]
\\
\dis
&-&\dis
\frac{1}{2}\Tr
\left[\left\{
 \left(\Gamma^{(2)}_k + \widehat R_k\right)^{-1}_{\bar \xi \xi}
-\left(\Gamma^{(2)}_k + \widehat R_k\right)^{-1}_{\xi \bar\xi}
\right\}
\left(\partial_t \widehat R_k\right)_{\bar \xi \xi} \right]
\eay
\ee
If one evaluates the RHS of this equation in terms of
position--space matrix elements then $\Gamma^{(2)}_k$ is defined
by a formula similar to \rf{2.25} and the integration
implied by ``Tr'' has to be interpreted as $\int d^dx \, \sqrt{\gb(x)}$.
The matrix elements in the ghost sector are defined in terms
of left derivatives, e.g.
\be\label{2.28}
\left(\left(\Gamma^{(2)}_k\right)_{\bar\xi \xi}
\right)_{\mu x}^{\;\;\;\;\;\; \nu y}
=
\frac{1}{\sqrt{\gb(x)}}\,
\frac{\delta}{\delta \xi^\mu(x)}\,
\frac{1}{\sqrt{\gb(y)}}\,
\frac{\delta \Gamma_k}{\delta \bar\xi_\nu(y)}
\ee
For any cutoff which is qualitatively similar to \rf{2.11a}
the traces on the RHS of eq.\rf{2.27} are well convergent, both
in the IR and the UV. By virtue of the factor $\partial_t \widehat R_k$,
the dominant contributions come from a narrow band of generalized
momenta centered around $k$. Large momenta are exponentially suppressed.

Solving the evolution equation \rf{2.27} with the appropriate
initial condition at the UV cutoff scale $\Lambda\to\infty$
is tantamount to computing the original functional integral \rf{2.2}.
In order to determine the correct initial value $\Gamma_\Lambda$ we
consider the following integral equation satisfied by $\Gamma_k$:
\be\label{2.29}
\begin{array}{c}
\dis
\exp\left\{-\Gamma_k[\bar h, \xi,\bar\xi ;\beta, \tau; \gb]\right\}=
\qquad\qquad
\qquad\qquad
\qquad\qquad
\qquad\qquad
\qquad
\\
\dis
\int \cd h \cd C \cd \bar C \,
\exp\Bigg[ -\widetilde S[h, C,\bar C ;\beta, \tau; \gb]
\qquad\qquad
\qquad\qquad
\qquad\qquad
\\
\dis
+\int d^dx \, \Big\{ \left(h_\mn -\bar h_\mn\right)
                \frac{\delta \Gamma_k}{\delta \bar h_\mn}
+\left(C^\mu-\xi^\mu\right)\frac{\delta \Gamma_k}{\delta \xi^\mu}
+\left(\bar C_\mu -\bar\xi_\mu\right)
\frac{\delta \Gamma_k}{\delta\bar\xi_\mu} \Big\} \Bigg]
\\
\dis
\cdot
\exp\left\{-\Delta_k S[h-\bar h, C-\xi, \bar C -\xi; \gb ]\right\}
\eay
\ee
Here
\be\label{2.30}
\widetilde S \equiv S+ S_{\rm gf} + S_{\rm gh}
-\int d^dx \, \sqrt{\gb}
\left\{
\beta^\mn \cl_C(\gb_\mn+h_\mn)+\tau_\mu C^\nu\partial_\nu C^\mu
\right\}
\ee
is expressed in terms of the ``microscopic'' fields
$(h,C,\bar C)$. Eq. \rf{2.29} obtains by inserting the definition
of $\Gamma_k$ into \rf{2.2} and using
\be\label{2.31}
\frac{\delta \widetilde\Gamma_k}{\delta \bar h_\mn}
=\sqrt{\gb} \, t^\mn
\qquad,\qquad
\frac{\delta \widetilde\Gamma_k}{\delta \bar\xi_\mu}
=-\sqrt{\gb}\, \sigma^\mu
\qquad,\qquad
\frac{\delta \widetilde\Gamma_k}{\delta \xi^\mu}
=-\sqrt{\gb}\, \bar\sigma_\mu
\ee
The crucial observation is that for $k\to \infty$ the last
exponential in \rf{2.29} becomes proportional to a
$\delta$--functional which equates the quantum fields
$(h,C,\bar C)$ to their classical counterparts:
\be\label{2.32}
e^{-\Delta_k S}\;
\lower +.5ex \hbox{$\widetilde{\scriptscriptstyle k\to \infty }$}
\;
\delta[h-\bar h]
\delta[C-\xi]
\delta[\bar C-\bar\xi]
\ee
As a consequence, the effective average action at the UV cutoff
reads\footnote{
Strictly speaking \rf{2.33} is correct only up to local terms which at
most change the bare parameters in $S$. Because the value of the bare
parameters has anyhow no physical significance we ignore these terms here.}
\be\label{2.33}
\bay
\dis
\Gamma_\Lambda[\bar h,\xi, \bar \xi; \beta, \tau; \gb]
&=&\dis
S[\gb+\bar h] + S_{\rm gf}[\bar h; \gb]
 + S_{\rm gh}[\bar h, \xi, \bar \xi; \gb]
\\
&&\dis
-\int d^dx\, \sqrt{\gb}
\left\{
\beta^\mn \cl_\xi (\gb_\mn+\bar h_\mn)
+\tau_\mu \xi^\nu\partial_\nu \xi^\mu\right\}
\eay
\ee
It is this action $\Gamma_\Lambda$ which has to be used as the
initial condition for the evolution equation.
We note that at the level of the functional $\bar\Gamma_k[g]$
eq.\rf{2.33} boils down to
\be\label{2.34}
\bar\Gamma_\Lambda[g_\mn]=S[g_\mn]
\ee
As $\Gamma^{(2)}_k$ involves derivatives with respect to $g_\mn$ at
fixed $\bar g_\mn$ it is clear that the evolution equation cannot
be formulated in terms of $\bar\Gamma_k$ alone, however.

Up to now we assumed that the fundamental action $S$ is positive
definite and the euclidean functional integral \rf{2.2} makes sense
as it stands. It is well known that this is not the case for the
Einstein--Hilbert action, for example, because the conformal factor
has a ``wrong sign'' kinetic term. Clearly it would be desirable
to have an evolution equation which can be applied in such
cases as well. It is quite remarkable therefore that the
renormalization group equation \rf{2.27}, with a properly chosen
cutoff, is well--defined even if $S$ and $\Gamma_k$ are not positive
definite. To see this, let us look at the first trace on the RHS of
\rf{2.27} and let us concentrate on the contribution of a fixed mode
$\phi$ contained in the metric. We assume that $\phi$ is an
eigenfunction of $\Gamma^{(2)}_k$ with eigenvalue $z_k p^2$ where $p^2$
is a positive eigenvalue of some covariant kinetic operator,
typically of the form $-\bar D^2$ + {\it R--terms}. For theories with
$S>0$, the wave function renormalization $z_k$ is positive
(at least for large $k$). In this case the general rule \cite{rw3,ah}
is to define the constant $\cz_k$ in the cutoff $R_k$, eq.\rf{2.11},
as $\cz_k=z_k$ because this guarantees that for the low--momentum
modes the effective inverse propagators $\Gamma^{(2)}+R_k$ becomes
$z_k(p^2+k^2)$, as it should be.

The important question is how $\cz_k$
should be chosen if $z_k$ is negative. If we continue
to use $\cz_k=z_k$, the evolution equation is perfectly well defined
because the inverse propagator $-\vert z_k\vert (p^2+k^2)$ never
vanishes, and the traces of \rf{2.27} are not suffering from any
IR problems.
In fact, if we write down the perturbative expansion for the
functional trace, for instance, it is clear that all propagators
are correctly cut off in the IR, and that loop momenta smaller than
$k$ are suppressed.
On the other hand, if we set $\cz_k=-z_k$, then $-\vert z_k\vert (p^2-k^2)$
introduces a spurios singularity at $p^2=k^2$, and the cutoff fails
to make the theory IR finite in this case.

At first sight the choice
$\cz_k=-z_k$ might have appeared more natural because only if
$\cz_k>0$ the factor
$\exp\left(-\Delta_k S\right)\sim \exp\left(-\int R_k \phi^2\right)$
is a damped exponential which suppresses the low momentum modes
in the usual way.
In this paper we shall nevertheless adopt the rule $\cz=z_k$
for either sign of $z_k$. We shall see that at least for the
Einstein--Hilbert truncation of section 4 the evolution equations
are well defined and consistent even though it is difficult
to give a meaning to the functional integral itself.
In the case $\cz_k=z_k<0$ the factor
$\exp\left(+\int \vert R_k\vert \phi^2\right)$ unavoidably becomes
a growing exponential and it might seem that this enhances rather
than suppresses the low momentum modes. However,
as suggested by the perturbative argument above,
this conclusion is too naive probably.
Moreover, if one invokes the usual prescription of rotating the
contour of integration over $\phi$ so that it is parallel to the
imaginary axis, both the kinetic term and the cutoff lead to damped
exponentials.

Furthermore, it is important to note that the constructions in this
section can be repeated for metrics on Lorentzian spacetimes.
Then one deals with oscillating exponentials $e^{iS}$, and for
arguments like the one leading to eq.\rf{2.32} one has to employ
the Riemann-Lebesgue lemma. Apart from the obvious substitutions
$\Gamma_k \to -i\Gamma_k$, $R_k\to -i R_k$, the evolution equation
remains unaltered. For $\cz_k=z_k$ it has all the desired features,
and $z_k<0$ seems not to pose any special problem.

\mysection{\protect\makebox[\textwidth][l]{Modified Ward Identities and}
           \protect\makebox[\textwidth][l]{Consistent Truncations}}
We mentioned already that the classical action plus
the gauge fixing and ghost terms are invariant under
the BRS transformations \rf{2.13}. Therefore the BRS variation
of the total action
$S_{\rm tot}\equiv S+S_{\rm gf}+S_{\rm gh}+\Delta_k S +S_{\rm sources}$
receives contributions only from the cutoff and the
source terms. If we apply a BRS transformation
to the integral defining $W_k$ and assume that the
measure is invariant we obtain
\be\label{3.1}
<\delta_\varepsilon S_{\rm sources} +\delta_\varepsilon\Delta_k S> =0
\ee
where
\be\label{3.2}
<\co> \equiv e^{-W_k}\int \cd h \cd C \cd \bar C\, \co \,e^{-S_{\rm tot}}
\ee
Our goal is to convert \rf{3.1} to a statement about
the average action $\Gamma_k$. Because the BRS transformation \rf{2.13}
is off--shell nilpotent when acting on $h_\mn$ and on $C^\mu$
(but not on $\bar C_\mu$) one has
\be\label{3.3}
\bay
\dis
\delta_\varepsilon S_{\rm sources}
&=&\dis
 -\varepsilon \kappa^{-2} \int d^dx \, \sqrt{\gb}
\Big\{
  t^\mn \cl_C (\gb_\mn + h_\mn)
\\
&&\dis
\qquad\qquad
\qquad\qquad
-\bar\sigma_\mu C^\nu \partial_\nu C^\mu
-\kappa \alpha^{-1} \sigma^\mu F_\mu(\gb,h)
\Big\}
\eay
\ee
If we take the expectation value of \rf{3.3} and express $W_k$
in terms of $\Gamma_k$ we find
\be\label{3.4}
<\delta_\varepsilon S_{\rm source}>
=
\frac{\epsilon}{\kappa^2}\int d^dx \,
\frac{1}{\sqrt{\gb(x)}}
\left\{
\frac{\delta\Gamma_k^\prime}{\delta \bar h_\mn}\,
\frac{\delta\Gamma_k^\prime}{\delta \beta^\mn}
+
\frac{\delta\Gamma_k^\prime}{\delta \xi^\mu}\,
\frac{\delta\Gamma_k^\prime}{\delta \tau_\mu}
\right\}
+
\frac{\varepsilon}{\kappa^2}\widetilde Y_k
\ee
with
\be\label{3.5}
\widetilde Y_k
\equiv
\int d^dx\, \left\{
\frac{1}{\sqrt{\gb}}
\left(
\frac{\delta\Delta_k S}{\delta \bar h_\mn}\,
\frac{\delta\Gamma_k^\prime}{\delta \beta^\mn}
+
\frac{\delta\Delta_k S}{\delta \xi^\mu}\,
\frac{\delta\Gamma_k^\prime}{\delta \tau_\mu}
\right)
-\sqrt{2} \frac{\kappa}{\alpha} \sqrt{\gb} F_\mu (\gb, \bar h)
R^{\rm gh}_k \xi^\mu
\right\}
\ee
Here we defined
\be\label{3.6}
\Gamma^\prime_k \equiv \Gamma_k - S_{\rm gf}[\bar h; \gb]
\ee
and we exploited the equation of motion
$<\delta S_{\rm tot}/\delta \bar C_\mu>=0$
which can be cast in the form
\be\label{3.7}
\left[
\frac{1}{\sqrt{\gb(x)}}\,
\frac{\delta}{\delta\bar\xi_\mu(x)}
-
\sqrt{2}\gb^\mn \cf^{\rho \sigma}_\nu
\frac{1}{\sqrt{\gb(x)}}\,
 \frac{\delta}{\delta\beta^{\rho\sigma}(x)}
\right]
\Gamma_k[\bar h,\xi, \bar \xi; \beta, \tau; \gb]=0
\ee
The variation of the cutoff terms gives rise to
\be\label{3.8}
<\delta_\varepsilon \Delta_k S> = -\frac{\varepsilon}{\kappa^2}
\left( Y_k +\widetilde Y_k\right)
\ee
with
\renewcommand{\arraystretch}{2.3}
\be\label{3.9}
\bay
\dis
Y_k
& \equiv &
\dis
\kappa^2 \Tr
\left[
  \left(R^{\rm grav}_k\right)^{\mn\rho\sigma}
  \left( \Gamma_k^{(2)}+\widehat R_k \right)^{-1}_{\bar h_{\rho\sigma}\varphi}
\,
\frac{\delta^2\Gamma_k}{\sqrt{\gb}\delta\varphi \, \sqrt{\gb}\delta \beta^\mn}
\right]
\\
&&\dis
-\sqrt{2}\Tr
\left[
  R^{\rm gh}_k
  \left( \Gamma_k^{(2)}+\widehat R_k \right)^{-1}_{\xi^\mu\varphi}
\,
\frac{\delta^2\Gamma_k}{\sqrt{\gb}\delta\varphi \, \sqrt{\gb}\delta \tau_\mu}
\right]
\\
&&\dis
+2\alpha^{-1}\kappa^2\Tr
\left[
  R^{\rm gh}_k \cf^{\rho\sigma}_\mu
  \left( \Gamma_k^{(2)}+\widehat R_k
\right)^{-1}_{\bar h_{\rho\sigma}\bar\xi_\mu}
\right]
\eay
\ee
where $\varphi\equiv\{ \bar h, \xi,\bar\xi\}$ is summed over.
From \rf{3.4} and \rf{3.8} we obtain the Ward identities in their
final form:
\be\label{3.10}
\int d^dx \, \frac{1}{\sqrt{\gb}} \,
\left\{
\frac{\delta\Gamma_k^\prime}{\delta \bar h_\mn}\,
\frac{\delta\Gamma_k^\prime}{\delta \beta^\mn}
+
\frac{\delta\Gamma_k^\prime}{\delta \xi^\mu}\,
\frac{\delta\Gamma_k^\prime}{\delta \gamma_\mu}
\right\}
=Y_k
\ee
Eq.\rf{3.10} has to be compared to the ordinary gravitational
Ward identities \cite{ojima} which are similar to \rf{3.10}
but with a vanishing RHS. In fact, the contribution $Y_k$ is
due to the cutoff and therefore it vanishes in the limit $k\to 0$
because $ R_k \sim k^2 \to 0$ in this limit.
Hence the standard effective action $\dis \lim_{k \to 0} \Gamma_k$
is guaranteed to obey its usual Ward identities, and BRS invariance
is restored for $k\to 0$.

Because the Ward identity \rf{3.10} is derived from the same functional
integral as the evolution equation, it is automatically satisfied
for the exact solution of the evolution equation. For approximate
solutions of the evolution equation their consistency with the
Ward identity is not guaranteed, and one may even use \rf{3.10} to
judge the quality of the approximation\cite{ehw,liou}.

The most important strategy for finding approximate
(but still nonperturbative) solutions to the evolution
equation is to truncate the space of action functionals.
Typically one works on a finite--dimensional subspace
parametrized by only a few generalized couplings.
As a first step towards such a truncation one can try
to neglect the evolution of the ghost action. This amounts
to making an ansatz of the following form:
\renewcommand{\arraystretch}{2.0}
\be\label{3.11}
\bay\dis
\Gamma_k[g,\gb ,\xi ,\bar\xi ;\beta ,\tau]
&=&\dis
\bar\Gamma_k[g]+\widehat\Gamma_k[g,\gb]
+S_{\rm gf}[g-\gb ;\gb] +S_{\rm gh}[g-\gb ,\xi,\bar\xi ;\gb]
\\
&&\dis
-\int d^dx\, \sqrt{\gb} \,
\left\{ \beta^\mn \cl_\xi g_\mn +\tau_\mu \xi^\nu\partial_\nu\xi^\mu
\right\}
\eay
\ee
In \rf{3.11} we pulled out the classical $S_{\rm gf}$ and $S_{\rm gh}$
from $\Gamma_k$, and also the coupling to the BRS variations
has the same form as in the bare action. The remaining functional
depends on both $g_\mn$ and $\bar g_\mn$.
It is further decomposed as $\bar\Gamma_k+\widehat\Gamma_k$ where
$\bar\Gamma_k$ is defined as in \rf{2.21} and $\widehat\Gamma_k$
contains the deviations for $\gb\ne g$. Hence by definition
\be\label{3.12}
\widehat\Gamma_k[g,g]=0
\ee
$\widehat\Gamma_k$ can be viewed as a quantum correction
the gauge fixing term which also vanishes for $\gb = g$.
The ansatz \rf{3.11} satisfies the initial condition \rf{2.33} if
\be\label{3.13}
\bar\Gamma_\Lambda = S
\qquad , \qquad
\widehat \Gamma_\Lambda =0
\ee
and it satisfies the quantum equation of motion \rf{3.7} exactly.
Eq. \rf{3.13} suggests to set $\widehat\Gamma_k=0$ for all $k$ in
a first approximation. In this case it can be checked that if
the ansatz \rf{3.11} is inserted into the Ward identity \rf{3.10}
its LHS vanishes identically.
Including $\widehat\Gamma_k$ the Ward identity assumes the form
\be\label{3.17}
\int d^dx \, \cl_\xi g_\mn\,
\frac{\delta \widehat\Gamma_k[g,\gb]}{\delta g_\mn(x)}
=-Y_k
\ee
We see that $\widehat\Gamma_k=0$ is a good approximation
provided we may neglect $Y_k$. The traces which define $Y_k$
amount to loop integrals, and if we think in terms of a
loop expansion $Y_k$ is certainly a higher loop effect
and may be neglected in a first approximation.
At the nonperturbative level one can still try to set
$\widehat\Gamma_k=0$ and investigate the consequences in
concrete examples.
In Yang--Mills theory the analogous truncation has led
to rather encouraging results already \cite{rw3,ah,topren}.
In the next section we shall perform an explicit
calculation in this approximation.

If one inserts the ansatz \rf{3.11} into the evolution
equation \rf{2.27} one finds the following equation for the
evolution of $\Gamma_k$ in the subspace spanned by the ansatz:
\be\label{3.18}
\bay
\dis
\partial_t\Gamma_k[g,\gb]
&=&\dis
\frac{1}{2}\Tr
\left[\left(
  \kappa^{-2}\Gamma^{(2)}_k[g,\gb] +R_k^{\rm grav}[\gb]
      \right)^{-1}
  \partial_t R^{\rm grav}_k[\gb]
\right]
\\
&&\dis
-\Tr\left[\left(
  -\cm[g,\gb]+R^{\rm gh}_k[\gb]
          \right)^{-1}
\partial_t R^{\rm gh}_k[\gb]
\right]
\eay
\ee
This equation is written down in terms of
\renewcommand{\arraystretch}{2.0}
\be\label{3.19}
\bay\dis
\Gamma_k[g,\gb] &=& \dis \Gamma_k[g,\gb,0,0;0,0]
\\
&=& \bar\Gamma_k[g]+S_{\rm gf}[g-\gb;\gb]
    +\widehat\Gamma_k[g,\gb]
\eay
\ee
$\Gamma^{(2)}_k$ is the Hessian of $\Gamma_k[g, \gb]$ with respect
to $g_\mn$ at fixed $\gb_\mn$. For the harmonic coordinate condition,
the classical kinetic term of the ghosts, $\cm$, is given by eq.\rf{2.9b}.

\mysection{The Einstein--Hilbert Truncation}
In this section we illustrate the use of eq.\rf{3.18}
by means of a simple example. At the UV scale $\Lambda$
we start from the classical Einstein--Hilbert action in
$d$ dimensions,
\be\label{4.1}
S=\frac{1}{16\pi\bar G} \int d^dx \,
\sqrt{g} \left\{ -R(g)+ 2\bar\lambda \right\},
\ee
and we evolve it down to smaller scales $k<\Lambda$.
For the time being we shall not try to send $\Lambda$ to infinity,
so the nonrenormalizability of the theory is not an issue here.
We are going to use a truncation which replaces in \rf{4.1}
the bare Newton constant $\bar G$ and the bare cosmological constant
$\bar \lambda$ by $k$--dependent functions
\be\label{4.2}
G_k\equiv \znk^{-1} \bar G
\ee
and $\bar\lambda_k$, respectively:
\be\label{4.3}
\bay
\dis
\Gamma_k[g,\gb] & = & \dis
2\kappa^2\znk \int d^dx \,\sqrt{g} \left\{ -R(g)+ 2\bar\lambda_k \right\}
\\
&&\dis +\kappa^2\znk \int d^dx \,\sqrt{\gb} \, \bar g^\mn
\left(\cf_\mu^{\alpha\beta} g_{\alpha\beta}\right)
\left(\cf_\nu^{\rho\sigma} g_{\rho\sigma}\right)
\eay
\ee
This ansatz is of the form \rf{3.19} with $\widehat\Gamma_k$ neglected
and the classical gauge fixing term given by \rf{2.4} with \rf{2.8},
\rf{2.9} and $\alpha=1/\znk$. (Note that
$\cf_\mu^{\alpha\beta} g_{\alpha\beta}=\cf_\mu^{\alpha\beta}
\bar h_{\alpha\beta}$ because $\bar D_\mu\gb_{\alpha\beta}=0$.)
In order to determine the functions $\znk$ and $\bar \lambda_k$ we have
to project the evolution equation on the space spanned by the
operators $\sqrt{g}$ and $\sqrt{g} R$. After having inserted the
ansatz into the evolution equation we may set $\gb_\mn=g_\mn$
so that the gauge fixing term in \rf{4.3} vanishes. The LHS of the
evolution equation reads then
\be\label{4.4}
\partial_t\Gamma_k[g,g]=2\kappa^2
\int d^dx \, \sqrt{g}
\left[-R(g)\partial_t \znk +2 \partial_t\left(\znk \bar\lambda_k\right)\right]
\ee
On the RHS of \rf{3.18} we have to perform a derivative expansion
and retain only the terms proportional to $\int\sqrt{g}$ and
$\int\sqrt{g}R$. Equating the result to \rf{4.4} we can read off the
system of ordinary differential equations for $\znk$ and $\bar\lambda_k$.
They have to be solved subject to  the initial conditions $Z_{N\Lambda}=1$
and $\bar\lambda_\Lambda = \bar\lambda$. In this manner the
renormalization group flow in the space of all action functionals is
projected onto the 2--dimensional subspace parametrized by
$\bar G$ and $\bar\lambda$.

In the evolution equation we need the second functional derivative
of $\Gamma_k[g,\gb]$ at fixed $\gb_\mn$. We expand
\be\label{4.5}
\Gamma_k[\gb+\bar h, \gb]
=\Gamma_k[\gb,\gb] + O(\bar h)
+\Gamma_k^{\rm quad}[\bar h; \gb] +O(\bar h^3)
\ee
and we find for the piece which is quadratic in $\bar h_\mn$:
\be\label{4.6}
\Gamma_k^{\rm quad}[\bar h; \gb]
=
\znk \kappa^2
\int d^dx \, \sqrt{\gb} \,\bar{h}_\mn
\left[ -K^\mn_{~~\rho\sigma} \bar D^2 + U^\mn_{~~\rho\sigma}\right]
\bar h^{\rho\sigma}
\ee
Here indices are raised and lowered with $\gb_\mn$, and the tensors
$K$ and $U$ are given by
\be\label{4.7}
K^\mn_{~~\rho\sigma}
=
\frac{1}{4}
\left[
 \delta^\mu_\rho   \delta^\nu_\sigma
+\delta^\mu_\sigma \delta^\nu_\rho
-\gb^\mn \gb_{\rho\sigma}
\right]
\ee
and
\be\label{4.8}
\bay\dis
U^\mn_{~~\rho\sigma}
&=&\dis
\frac{1}{4}
\left[
 \delta^\mu_\rho   \delta^\nu_\sigma
+\delta^\mu_\sigma \delta^\nu_\rho
-\gb^\mn \gb_{\rho\sigma}
\right]
\left(\bar R- 2\bar\lambda_k\right)
+
\frac{1}{2}
\left[\gb^\mn\bar R_{\rho\sigma} +\gb_{\rho\sigma}\bar R^\mn\right]
\\
&-&\dis
\frac{1}{4}
\left[
  \delta^\mu_\rho \bar R^\nu_{~\sigma}
 +\delta^\mu_\sigma \bar R^\nu_{~\rho}
 +\delta^\nu_\rho \bar R^\mu_{~\sigma}
 +\delta^\nu_\sigma \bar R^\mu_{~\rho}
\right]
-
\frac{1}{2}
\left[
  \bar R^{\nu~\mu}_{~\rho~\sigma}
 +\bar R^{\nu~\mu}_{~\sigma~\rho}
\right]
\eay
\ee
In eq.\rf{4.8} all geometrical quantities are constructed
from the background metric.\footnote{
We use the conventions
$R^\sigma_{~\rho\mn}=-\partial_\nu \Gamma^\sigma_{\mu\rho}+\dots$,
$R_\mn =R^\sigma_{~\mu\sigma\nu}$ and $R=g^\mn R_\mn$.
}
In order to partially diagonalize the quadratic form \rf{4.6} we
write $\bar h_\mn$ as the sum of a traceless tensor
$\widehat h_\mn$ and a trace part involving $\phi\equiv\gb^\mn \bar h_\mn$:
\be\label{4.9}
\hb_\mn=\wh h_\mn+d^{-1}\gb_\mn\phi
\quad ,\quad
\gb^\mn\wh h_\mn=0
\ee
As a consequence, eq.\rf{4.6} becomes
\be\label{4.10}
\bay\dis
\Gamma_k^{\rm quad}[\hb;\gb]
&=&\dis
\znk \kappa^2 \int d^dx \, \sqrt{\gb}
\Bigg\{
  \frac{1}{2}\,\wh h_\mn
    \left[-\bar D^2-2\bar\lambda_k+\rb\right] \wh h^\mn
\\
&&\dis
\qquad
\qquad
\quad
-\left(\frac{d-2}{4d}\right)
\phi\left[-\bar D^2-2\bar\lambda_k+\frac{d-4}{d}\rb\right]\phi
\\
&&\dis
\qquad
\qquad
\quad
-\rb_\mn\wh h^{\nu\rho} \wh h^\mu_{~\rho}
+\rb_{\alpha\beta\nu\mu} \wh h^{\beta\nu} \wh h^{\alpha\mu}
+
\frac{d-4}{d}\phi \rb_\mn \wh h^\mn
\Bigg\}
\eay
\ee
The equations for $\znk$ and $\bar\lambda_k$ obtain by comparing
the coefficients of $\int\sqrt{g}$ and $\int \sqrt{g}R$ on both sides
of the evolution equation at $\gb_\mn=g_\mn$. For this purpose
we may insert an arbitrary family of metrics $g_\mn$ which is general
enough to identify the terms $\int\sqrt{g}$ and $\int\sqrt{g}R$ and to
distinguish them from higher order terms in the derivative expansion,
such as $\int\sqrt{g}R^2$ or $\int\sqrt{g}R^\mn D_\mu D_\nu R$,
for instance. We exploit this freedom by assuming that $\gb_\mn$
corresponds to a maximally symmetric space, i.e., that
\be\label{4.11}
\bay
\rb_{\mn\rho\sigma}
&=&\dis
\frac{1}{d(d-1)}
\left[
  \gb_{\mu\rho}\gb_{\nu\sigma}
 -\gb_{\mu\sigma}\gb_{\nu\rho}
\right]\rb
\\
\rb_\mn
&=&\dis
\frac{1}{d}\gb_\mn\rb
\eay
\ee
From now on the curvature scalar $\rb$ parametrizes the family of
metrics inserted, and it should be regarded as an externally
prescribed number rather than a functional of the metric. For a
maximally symmetric background the quadratic action boils down to
\be\label{4.12}
\bay\dis
\Gamma_k^{\rm quad}[\hb;\gb]
&=&\dis
\frac{1}{2}\znk\kappa^2
\int d^dx\,
\Bigg\{
  \wh h_\mn
  \left[-\bar D^2-2\bar\lambda_k+C_T\rb\right]\wh h^\mn
\\
&&\dis
\qquad
\qquad
\qquad
-\left(\frac{d-2}{2d}\right)\phi
  \left[-\bar D^2-2\bar\lambda_k+C_S\rb\right]\phi
\Bigg\}
\eay
\ee
with
\be\label{4.13}
C_T\equiv\frac{d(d-3)+4}{d(d-1)}
\qquad , \qquad
C_S \equiv \frac{d-4}{d}
\ee

Before continuing we have to specify the precise form
of the cutoff operators $R_k^{\rm grav}$ and $R_k^{\rm gh}$
to be used in the evolution equation \rf{3.18}. Both of them
have the structure \rf{2.11} whereby $\cz_k$ should be adjusted
in such a way that for every low--momentum mode the cutoff
combines with the kinetic term of this mode to $-\bar D^2+k^2$
times a constant. Looking at \rf{4.12} we see that the respective
kinetic terms for $\wh h_\mn$ and $\phi$ differ by a factor of
$-(d-2)/2d$. This suggests the following choice:
\be\label{4.14}
\left(\cz_k^{\rm grav}\right)^{\mn\rho\sigma}
=
\left[
  \left(I-P_\phi\right)^{\mn\rho\sigma}
 -\frac{d-2}{2d} P_\phi^{\mn\rho\sigma}
\right]
\znk
\ee
Here
\be\label{4.15}
\left(P_\phi\right)_\mn^{~~\rho\sigma}
=d^{-1} \gb_\mn\gb^{\rho\sigma}
\ee
is the projector on the trace part of the metric.
For the traceless tensor \rf{4.14} coincides with \rf{2.10a} for
$Z_k^{\rm grav}=\znk$, and for $\phi$ the different relative
normalization is taken into account. Thus we obtain in the $\wh h$
and the $\phi$--sector, respectively:
\be\label{4.16}
\bay\dis
\left(
  \kappa^{-2}\Gamma_k^{(2)}[g,g]+R_k^{\rm grav}
\right)_{\wh h \wh h}
&=&\dis
\znk
\left[-D^2+k^2 R^{(0)}(-D^2/k^2)-2\bar\lambda_k+C_T R\right],
\\
\dis
\left(
  \kappa^{-2}\Gamma_k^{(2)}[g,g]+R_k^{\rm grav}
\right)_{\phi\phi}
&=&\dis
-\frac{d-2}{2d}
\znk
\left[-D^2+k^2 R^{(0)}(-D^2/k^2)-2\bar\lambda_k+C_S R\right]
\eay
\ee
From now on we may set $\gb=g$ and we omit the bars from the
metric and the curvature.

The last missing ingredient for the evolution equation is the
Faddeev--Popov operator. From \rf{2.9b} one obtains at $\gb=g$
\be\label{4.17}
\cm[g,g]^\mu_{~\nu}=\delta^\mu_\nu D^2 +R^\mu_{~\nu}
=-\delta^\mu_\nu\left[-D^2+C_V R\right]
\ee
with
\be\label{4.18}
C_V\equiv -\frac{1}{d}
\ee
In the second part of \rf{4.17} we used \rf{4.11} for
a maximally symmetric background. Since we did not take
into account any renormalization effects in the ghost action
we set $Z_k^{\rm gh}\equiv1$ in $R_k^{\rm gh}$ and obtain
\be\label{4.19}
-\cm + R_k^{\rm gh} = - D^2 +k^2 R^{(0)}(-D^2/k^2)+C_V R
\ee

Let us write $\cs_k(R)$ for the RHS of the renormalization
group equation \rf{3.18} with $\gb=g$. Inserting \rf{4.16} and
\rf{4.19} there we arrive at
\be\label{4.20}
\bay\dis
\cs_k(R)
&=&\dis
\Tr_T\left[\cn(\ca+C_TR)^{-1}\right]
+
\Tr_S\left[\cn(\ca+C_SR)^{-1}\right]
\\
&-&\dis
2\Tr_V\left[\cn_0(\ca_0+C_VR)^{-1}\right]
\eay
\ee
with
\be\label{4.21}
\bay\dis
\ca
&\equiv&\dis
-D^2+k^2R^{(0)}(-D^2/k^2)-2\bar\lambda_k
\\
\dis\cn
&\equiv&\dis
(2\znk)^{-1}\partial_t
\left[\znk k^2 R^{(0)}(-D^2/k^2)\right]
\\
&=&\dis
\left[1-\frac{1}{2}\eta_N(k)\right]
k^2 R^{(0)}(-D^2/k^2)
+D^2 R^{(0)\,\prime}(-D^2/k^2)
\eay
\ee
where a prime denotes the derivative with respect to the argument and
\be\label{4.22}
\eta_N(k)\equiv -\partial_t\ln\znk
\ee
is the anomalous dimension of the operator $\sqrt{g}R$.
The operators $\cn_0$ and $\ca_0$ are defined similarly
to \rf{4.21} but with $\lambda=0$ and $\znk=1$, i.e.,
$\eta_N(k)=0$. Eq.\rf{4.20} involves traces of functions of the
covariant Laplacian $D^2\equiv g^\mn D_\mu D_\nu$ acting on
traceless symmetric tensors (``$T$''), scalars (``$S$'')
and vectors (``$V$''). Because we need  only the zeroth and the
first order in the curvature scalar we can expand
\be\label{4.23}
\kern -.9em
\bay\dis
\cs_k(R)
&=&\dis
 \Tr_T\left[\cn\ca^{-1}\right]
+\Tr_S\left[\cn\ca^{-1}\right]
-2\Tr_V\left[\cn_0\ca^{-1}_0\right]
\\
&-&\dis
\!\!
R\left(
 C_T  \Tr_T\!\!\left[\cn\ca^{-2}\right]
+C_S  \Tr_S\!\!\left[\cn\ca^{-2}\right]
-2 C_V  \Tr_V\!\!\left[\cn_0\ca^{-2}_0\right]
\right)
\!+\!O(R^2)
\eay
\ee
The traces in \rf{4.23} can be evaluated by taking advantage of the
heat kernel expansion
\be\label{4.23a}
\Tr\left[e^{-isD^2}\right]
=
\left(\frac{i}{4\pi s}\right)^{d/2}
\tr(I)
\int d^dx\, \sqrt{g}
\left\{
  1-\frac{1}{6}isR+O(R^2)
\right\}
\ee
Here $I$ denotes the unit matrix of the space of fields on which
$D^2$ acts. Hence $\tr(I)$ is the number of independent field
components and in particular
\renewcommand{\arraystretch}{1.2}
\be\label{4.24}
\bay
\tr_S(I)&=&1
\\
\tr_V(I)&=&d
\\
\tr_T(I)&=&\dis \frac{1}{2}(d-1)(d+2)
\eay
\ee
Considering an arbitrary function $W$ with a Fourier transform
$\widetilde W$, the expansion of the trace
\be\label{4.24a}
\Tr[W(-D^2)]=\int_{-\infty}^{\infty} ds \,
\widetilde W(s)\, \Tr\left[e^{-isD^2}\right]
\ee
is given by
\renewcommand{\arraystretch}{2.0}
\be\label{4.25}
\bay\dis
\Tr[W(-D^2)]
&=&\dis
(4\pi)^{-d/2} \tr(I)
\Bigg\{
  Q_{d/2}[W] \int d^dx \, \sqrt{g}
\\
&&\dis
\qquad\qquad
\qquad
+\frac{1}{6} Q_{d/2-1}[W] \int d^dx\, \sqrt{g}R
+O(R^2)
\Bigg\}
\eay
\ee
with
\be\label{4.26}
Q_n[W]\equiv
\int_{-\infty}^{\infty} ds\, (-is)^n \widetilde W(s)
\ee
Reexpressing \rf{4.26} in terms of $W$ leads to the Mellin
transform ($n>0$)
\be\label{4.27}
\bay\dis
Q_0[W] &=& \dis W(0)
\\
\dis
Q_n[W] &=&\dis
 \frac{1}{\Gamma(n)} \int_0^\infty dz\, z^{n-1} W(z)
\eay
\ee
The next step is to use \rf{4.25} in order to evaluate
\rf{4.23} and to combine $\cs(R)$ with the LHS of the evolution
equation, eq.\rf{4.4}. From the coefficients of $\int\sqrt{g}$
we can read off the following equation
\be\label{4.28}
\kern -.6em
\bay\dis
\partial_t\left(\znk \bar\lambda_k\right)
&=&\dis
(4\kappa^2)^{-1} (4\pi)^{-d/2}
\Bigg\{
  \tr_T(I) Q_{d/2}[\cn/\ca]
\\
&&\dis
\qquad
\qquad
\qquad
+\tr_S(I) Q_{d/2}[\cn/\ca]
-2\tr_V(I) Q_{d/2} [\cn_0/\ca_0]
\Bigg\}
\eay
\ee
Likewise $\int\sqrt{g} R$ gives rise to
\be\label{4.29}
\kern -.6em
\bay\dis
\partial_t\znk
&=&\dis
-(12\kappa^2)^{-1}(4\pi)^{-d/2}
\Bigg[
  \tr_T(I)\left\{Q_{d/2-1}[\cn/\ca]-6C_T Q_{d/2} [\cn/\ca^2] \right\}
\\
&&\dis
\qquad\qquad
\qquad\qquad
  +\tr_S(I)\left\{Q_{d/2-1}[\cn/\ca]-6C_S Q_{d/2} [\cn/\ca^2] \right\}
\\
&&\dis
\qquad\qquad
\qquad
  -2\tr_V(I)\left\{Q_{d/2-1}[\cn_0/\ca_0]-6C_V
               Q_{d/2} [\cn_0/\ca_0^2] \right\}
\Bigg]
\eay
\ee
In \rf{4.28} and \rf{4.29}, $\cn$ and $\ca$ are considered
c--number functions of $z$ which replaces $-D^2$ in \rf{4.21}. For
every cutoff $R^{(0)}$ we define the functions $(p=1,2,\dots)$
\be\label{4.30}
\bay\dis
\Phi^p_n(w)
&=&\dis
\frac{1}{\Gamma(n)}\int_0^\infty dz\,
z^{n-1}
\frac{R^{(0)}(z)-y R^{(0)\,\prime}(z)}{[z+R^{(0)}(z)+w]^p}
\\
\widetilde\Phi^p_n(w)
&=&\dis
\frac{1}{\Gamma(n)}\int_0^\infty dz\,
z^{n-1}
\frac{R^{(0)}(z)}{[z+R^{(0)}(z)+w]^p}
\eay
\ee
for $n>0$, and\footnote{
Actually eq.\rf{4.31} follows from \rf{4.30} in the limit $n\searrow 0$.}
\be\label{4.31}
\Phi^p_0(w)=\widetilde\Phi^p_0(w)=(1+w)^{-p}
\ee
In terms of the $\Phi$'s, eq.\rf{4.28} assumes the form
\be\label{4.32}
\bay\dis
\partial_t\left(\znk \bar\lambda_k\right)
&=&\dis
(16\kappa^2)^{-1}(4\pi)^{-d/2} k^d
\bigg[
  2d(d+1) \Phi^1_{d/2}(-2\bar\lambda_k/k^2)-8d\Phi^1_{d/2}(0)
\\
&&\dis
\qquad\qquad
\qquad\qquad
  -d(d+1)\eta_N \widetilde\Phi^1_{d/2}(-2\bar\lambda_k/k^2)
\bigg]
\eay
\ee
and \rf{4.29} becomes
\be\label{4.33}
\bay\dis
\partial_t\znk
&=&\dis
-(24\kappa^2)^{-1}(4\pi)^{-d/2}k^{d-2}\cdot
\\
&&\dis
\qquad
\cdot
\Bigg[
 d(d+1)\left\{ \Phi^1_{d/2-1}(-2\bar\lambda_k/k^2)
             -\frac{1}{2}\eta_N
             \widetilde\Phi^1_{d/2-1}(-2\bar\lambda_k/k^2)
       \right\}
\\
&&\dis
\qquad
 -6d(d-1)\left\{ \Phi^2_{d/2}(-2\bar\lambda_k/k^2)
             -\frac{1}{2}\eta_N
             \widetilde\Phi^2_{d/2}(-2\bar\lambda_k/k^2)
         \right\}
\\
&&\dis
\qquad
-4d\Phi^1_{d/2-1}(0)-24\Phi^2_{d/2}(0)
\Bigg]
\eay
\ee
Let us introduce the dimensionless, renormalized Newton constant
\be\label{4.34}
g_k\equiv k^{d-2}G_k \equiv k^{d-2} \znk^{-1} \bar G
\ee
and the dimensionless cosmological constant
\be\label{4.35}
\lambda_k\equiv k^{-2}\bar\lambda_k
\ee
Here $G_k\equiv \znk^{-1}\bar G$ is the dimensionful renormalized
Newton constant at scale $k$. The evolution of $g_k$ is governed
by the equation
\be\label{4.36}
\partial_t g_k = \left[d-2+ \eta_N(k)\right]g_k
\ee
From \rf{4.33} we obtain for the anomalous dimension $\eta_N(k)$:
\be\label{4.37}
\eta_N(k)=g_k B_1(\lambda_k) +\eta_N(k) g_k B_2(\lambda_k)
\ee
with
\be\label{4.38}
\bay\dis
B_1(\lambda_k)
&\equiv&\dis
\frac{1}{3}(4\pi)^{1-d/2}
\Bigg[
 d(d+1) \Phi^1_{d/2-1}(-2\lambda_k)
-6d(d-1)\Phi^2_{d/2}(-2\lambda_k)
\\
&&\dis
\qquad\qquad
\quad
-4d\Phi^1_{d/2-1}(0)-24\Phi^2_{d/2}(0)\Bigg]
\\
B_2(\lambda_k)
&\equiv&\dis
-\frac{1}{6}(4\pi)^{1-d/2}
\left[
 d(d+1) \widetilde\Phi^1_{d/2-1}(-2\lambda_k)
-6d(d-1)\widetilde\Phi^2_{d/2}(-2\lambda_k)
\right]
\eay
\ee
We can solve \rf{4.37} for the anomalous dimension in terms of
$g_k$ and $\lambda_k$:
\be\label{4.39}
\eta_N=\frac{g_kB_1(\lambda_k)}{1-g_kB_2(\lambda_k)}
\ee
The scale derivative of $\lambda_k$ is related to \rf{4.32}
according to
\be\label{4.40}
\partial_t\lambda_k=-(2-\eta_N)\lambda_k
+32\pi g_k \kappa^2 k^{-d}
\partial_t(\znk \bar\lambda_k)
\ee
so that
\be\label{4.41}
\bay\dis
\partial_t\lambda_k
&=&\dis
-(2-\eta_N)\lambda_k+\frac{1}{2}g_k(4\pi)^{1-d/2}\cdot
\\
&&\dis
\quad
\cdot\left[
2d(d+1) \Phi^1_{d/2}(-2\lambda_k)-8d \Phi^1_{d/2}(0)
-d(d+1)\eta_N\widetilde\Phi^1_{d/2}(-2\lambda_k)\right]
\eay
\ee
Eqs.\rf{4.36} and \rf{4.41} with \rf{4.39} is the set of
differential equations we wanted to derive. Once the initial
values $g_\Lambda$ and $\lambda_\Lambda$ are given, it determines
the value of the running Newton's constant and cosmological constant at any
scale $k\le \Lambda$. Although they were derived from a relatively
simple truncation, the above evolution equations encapsulate
nonperturbative effects which go beyond a simple one--loop
calculation. This is particularly obvious if one expands for instance
\rf{4.39} for small values of $g_k$:
\be\label{4.42}
\eta_N=g_k B_1(\lambda_k)
\left[1+g_k B_2(\lambda_k)+g_k^2B_2^2(\lambda_k)+\cdots\right]
\ee
We observe that $\eta_N$ receives contributions from
arbitrarily high orders of perturbation theory.

\mysection{\protect\makebox[\textwidth][l]{Running Newton's Constant and}
           \protect\makebox[\textwidth][l]{Cosmological Constant}}
\subsection{Near two dimensions}
In $d=2$ dimensions $\int\sqrt{g}R$ is a topological invariant
proportional to the Euler number and the quantum theory
under consideration has at most finitely many (topological)
degrees of freedom. In $d=2+\varepsilon$ dimensions, on the other
hand, one finds a dynamically nontrivial theory with a nonzero
$\beta$--function for $g_k$ \cite{eps,wein,kawai}:
\be\label{4.43}
\partial_t g_k=\left[\varepsilon +\eta_N\right] g_k
\ee
Gravity in $2+\varepsilon$ dimensions provides an interesting
laboratory for a first test of the evolution equation because
here the conformal factor of the metric can have both a conventional
($\varepsilon<0$) and a ``wrong--sign'' ($\varepsilon>0$) kinetic
term, see eq.\rf{4.12}.

The anomalous dimension has a power series expansion
\be\label{4.44}
\eta_N=\eta_N^{(0)}+\eta_N^{(1)}\varepsilon
+\eta_N^{(2)}\varepsilon^2+\dots
\ee
and therefore
\be\label{4.45}
\partial_t g_k=
\left[
  \left(1+\eta_N^{(1)}\right)\varepsilon +\eta_N^{(0)}
\right]g_k+O(\varepsilon^2)
\ee
Expanding the functions \rf{4.38} as
$B_{1,2}=B_{1,2}^{(0)}+B_{1,2}^{(1)}\varepsilon+\dots$
one has
\renewcommand{\arraystretch}{2.3}
\be\label{4.46}
\bay\dis
\eta_N^{(0)}&=&\dis
\frac{g_k B_1^{(0)}}{1-g_k B_2^{(0)}}
\\
\eta_N^{(1)} &=&\dis
 \frac{g_kB_1^{(1)}}{1-g_k B_2^{(0)}}
+\frac{g_k^2B_1^{(0)}B_2^{(1)}}{\left(1-g_k B_2^{(0)}\right)^2}
\eay
\ee
The lowest order terms are
\renewcommand{\arraystretch}{2.0}
\be\label{4.47}
\bay\dis
B_1^{(0)}(\lambda_k)
&=&\dis
2(1-2\lambda_k)^{-1}-4\Phi_1^2(-2\lambda_k)-\frac{32}{3}
\\
B_2^{(0)}(\lambda_k)
&=&\dis
2\widetilde\Phi_1^2(-2\lambda_k)-(1-2\lambda_k)^{-1}
\eay
\ee
We remark that for vanishing cosmological constant, $B_1^{(0)}$ is
a universal quantity, i.e., it does not depend on the precise form
of $R^{(0)}$:
\be\label{4.48}
B_1^{(0)}(0)=-\frac{38}{3}
\ee
The reason is that the integrand in the integral representation
of $\Phi_1^2(0)$ equals the derivative of $z(z+R^{(0)}(z))^{-1}$;
hence it is sufficient to know that $R^{(0)}$ is bounded
everywhere in order to establish that
\be\label{4.49}
\Phi_1^2(0)=1
\ee
Unlike $\Phi_1^2(0)$, $\widetilde\Phi_1^2(\lambda_k)$ is sensitive
to the shape of $R^{(0)}$ even for $\lambda_k=0$. In order to be
more explicit
we evaluate \rf{4.47} at $\lambda\ne0$ for the constant cutoff
function $R^{(0)}(z)=1$. Though it does not vanish for $z\to\infty$,
it yields at least qualitatively correct results \cite{ah, topren}
as long as it does not introduce UV divergences into
the integral under consideration. For $\Phi_1^2$ and $\widetilde\Phi_1^2$
this is not the case and one finds
\be\label{4.50}
\Phi_1^2(w)=\widetilde\Phi_1^2(w)=(1+w)^{-1}
\ee
so that
\be\label{4.51}
\bay\dis
B_1^{(0)}(\lambda_k)
&=&\dis
-2(1-2\lambda_k)^{-1}-\frac{32}{3}
\\
B_2^{(0)}(\lambda_k)
&=&\dis
(1-2\lambda_k)^{-1}
\eay
\ee
As a consequence, we obtain the following answer for the anomalous
dimension:
\be\label{4.52}
\eta_N^{(0)}=-\frac{38}{3}g_k
\frac{1-\frac{32}{19}\lambda_k}{1-g_k-2\lambda_k}
\ee
Eq.\rf{4.52} improves on earlier results in refs.\cite{eps,wein,kawai}.
It takes into account partially resummed higher loop effects
(higher powers of $g_k$) and it includes the effect of the  running
cosmological constant.

One of the interesting features of Einstein--Hilbert gravity in
$2+\varepsilon$ dimensions is that the evolution of Newton's
constant is governed by a fixed point $g_*$ at which the
$\beta$--function \rf{4.45} vanishes. To lowest order in
$\varepsilon$ it is given by
\be\label{4.53}
g_*=-\varepsilon B_1^{(0)}(\lambda_k)^{-1}
\ee
The $\lambda$--dependence of $g_*$ is non--universal.
For $R^{(0)}=1$ we obtain
\be\label{4.54}
g_*=\frac{3}{38}\,\varepsilon\,
\frac{1-2\lambda_k}{1-\frac{32}{19}\lambda_k}
\ee
Eq.\rf{4.54} is reliable for $\lambda_k\ll 1$. In this regime the
fixed point $g_*$ is UV stable if $\varepsilon>0$ and it is
IR stable for $\varepsilon<0$. For $\varepsilon>0$ and
$\lambda_k\equiv0$ this fixed point was discussed by Weinberg \cite{wein}
in the context of the asymptotic safety scenario for quantum gravity.
Our result for the dependence of $g_k$ on the cosmological constant
can only be obtained in a framework with a proper infrared
regularization because we are investigating the influence of the
relevant dimension--two operator on a marginal coupling.
(In a sense, the r\^ole played by the running cosmological constant
is similar to the quadratic mass renormalization in four dimensional
scalar theories.)
For $\varepsilon>0$ the theory is asymptotically free. Near the fixed
point the dimensionful Newton constant
$G_k =g_*/k^\varepsilon$ vanishes for $k\to\infty$.

The evolution of $\lambda_k$ itself is governed by eq.\rf{4.41}.
For $g_k\approx g_*$ where  $g_k$ and $\eta_N$ are of order
$\varepsilon$, one finds that also the $\beta$--function of
$\lambda$ has a zero of order $\varepsilon$:
\be\label{4.54a}
\lambda_*=-\frac{3}{38}\Phi_1^1(0)\varepsilon
\ee
This fixed point of the $\lambda$--evolution is UV stable
for either sign of $\varepsilon$. We conclude that to first
order in $\varepsilon$ and for $\varepsilon>0$ the combined
$(\lambda,g)$--system has an UV stable fixed point
given by \rf{4.54a} together with $g_*=(3/38)\varepsilon$.

\subsection{Four dimensions}
In $d=4$ dimensions, the running of Newton's constant
is governed by the following functions of the cosmological
constant:
\begin{samepage}
\begin{eqnarray}
\label{4.55a}
B_1(\lambda)
&=&
-\frac{1}{3\pi}
\left[18\Phi^2_2(-2\lambda)-5\Phi^1_1(-2\lambda)
    +6\Phi^2_2(0)+4\Phi^1_1(0)\right]
\\[10pt]
\label{4.55b}
B_2(\lambda)
&=&
\frac{1}{6\pi}
\left[18\widetilde\Phi^2_2(-2\lambda)-5\widetilde\Phi^1_1(-2\lambda)
\right]
\ea
\end{samepage}
The dimensionful quantity $G_k$ evolves according to
\be\label{4.56}
\partial_t G_k =\eta_N G_k
\ee
with the anomalous dimension given by \rf{4.39}. In order to get a
feeling for the behavior of $G_k$, let us restrict our attention to
the lowest order in $g_k$ which amounts to keeping only the first
nontrivial correction of the expansion in $\bar Gk^2$. Then
$\eta_N=B_1(\lambda_k)g_k+\dots$,
or with $g_k=k^2G_k= k^2 \bar G+O(\bar G^2)$,
\be\label{4.57}
\eta_N=B_1(\lambda_k) \bar G k^2 +O(\bar G^2)
\ee
First we consider the case where the cosmological constant
is much smaller than $k^2$. Then we may approximate $\lambda_k\approx 0$
in \rf{4.57}, and \rf{4.56} has the solution
\be\label{4.58}
G_k=G_0
\left[1-\omega \,\bar G k^2 +O(\bar G^2 k^4)\right]
\ee
Here
\be\label{4.59}
\omega\equiv -\frac{1}{2}B_1(0)
=\frac{1}{6\pi}\left[24\Phi^2_2(0)-\Phi^1_1(0)\right]
\ee
is a pure number, which depends on the function $R^{(0)}$, however.
For the exponential cutoff \rf{2.11a} we have $\Phi^1_1(0)=\pi^2/6$
and $\Phi^2_2(0)=1$, so that
\be\label{4.60}
\omega=\frac{4}{\pi}\left(1-\frac{\pi^2}{144}\right)>0
\ee
For different cutoff functions the numerical value of $\omega$ will be
slightly different but it will still be positive. Therefore
eq.\rf{4.58} tells us that Newton's constant decreases as $k^2$ increases;
it is small in the UV and grows larger as we evolve it towards the
infrared. The sign of this effect is the same as for the non--abelian
gauge coupling in Yang--Mills theory and it is opposite to the one
in QED. The main difference is that $G_k$ depends quadratically on
$k$ while,
to lowest order, the gauge coupling in Yang--Mills theory runs
only logarithmically. We see that gravity is ``antiscreening'' in
the sense that at large distances Newton's constant is larger than
at small distances. This confirms the intuitive picture
that the gravitational charge (mass) is not screened by
quantum fluctuations but rather receives an additional positive
contribution from the virtual particles surrounding it.

Let us consider a gravitational (thought) experiment which
involves a typical length scale $r$, the distance of two heavy
test particles, for instance. If $r\equiv k^{-1}$ acts as the
effective IR cutoff scale, eq.\rf{4.58} suggests the following
form of a distance--dependent Newton's constant (with factors of
$\hbar$ and $c$ restored):
\be\label{4.61}
G(r)=G(\infty)
\left[1-\omega\,\frac{\bar G \hbar}{r^2 c^3}
+O\left(\frac{1}{r^4}\right)
\right]
\ee
We expect\footnote{Recall that in QED the analogous substitution
$e^2/r\to e^2(r^{-1})/r$ correctly reproduces the leading term
of the Uehling potential if the one--loop formula for the
running coupling $e^2(\mu)$ is used \cite{dr1}.}
that, to leading order in $1/r$, the quantum corrected static
Newtonian potential of two test masses should be closely
related to $V(r)=-G(r)m_1m_2/r$.
It is interesting to compare \rf{4.61} to what is actually
obtained by a diagrammatic calculation of the lowest order
correction to the potential. Recently Donoghue \cite{don}
has pointed out that quantized Einstein gravity makes a well
defined prediction for this quantity which is unaffected by the
nonrenormalizability of the theory. One finds a result of the
form
\be\label{4.62}
V(r)=-G\frac{m_1m_2}{r}
\left[
  1-\frac{G(m_1+m_2)}{2c^2r}-\widetilde\omega \,\frac{G\hbar}{r^2c^3}
\right]
\ee
The term proportional to $(m_1+m_2)/r$ is a kinematic effect
of classical general relativity; it is independent of $\hbar$
and is not related to the $\beta$--function of $G_k$ therefore.
However, the last term in \rf{4.62}, proportional to $G\hbar/r^2$,
has precisely the same structure as \rf{4.61}. The most recent
calculation of $\widetilde\omega$ was performed in ref.\cite{hamli}
with the result
\be\label{4.63}
\widetilde\omega =\frac{118}{15\pi}>0
\ee
This number has the same sign and is of the same order of
magnitude as the value found originally in ref.\cite{don},
but there is no precise agreement yet. In ref.\cite{muz},
$\widetilde\omega$ was calculated using different methods
\cite{moda,hawi} and a negative value was found; this would
correspond to ``screening'' rather than ``antiscreening''.
Possible reasons for this discrepancy were discussed in
ref.\cite{hamli}. While the issue is not fully settled yet,
it is believed that by correctly identifying and evaluating
the set of relevant Feynman diagrams, quantum Einstein gravity
gives rise to an unambiguous value for $\widetilde\omega$.
From our investigation of the renormalization group flow we
expect this value to be positive.

One can use the full nonperturbative information contained
in \rf{4.39} in order to extend the domain of validity of our
result towards larger values of $g_k$ or smaller distances $r$.
This would involve a numerical solution of eq.\rf{4.36} on
which we shall not embark at this point.

In our approach we can study the influence of the cosmological
constant on the running of $G_k$. It is an interesting question,
for instance, whether a large $\lambda_k$ can destroy the
antiscreening character of the gravitational interaction
$(\eta_N<0)$. Let us look at \rf{4.57} with $B_1(\lambda_k)$
given in \rf{4.55a}.
If there exists a regime with $\eta_N>0$ (screening) then
$B_1(\lambda)$ must be positive there. This can only happen
if the term $5\Phi^1_1(-2\lambda_k)$ in the brackets on the
RHS of \rf{4.55a} is larger then the sum of the other terms
because the $\Phi$'s are always positive.
However, $\Phi^p_n(w)$ decreases for increasing $w$ and finally
vanishes for $w\to \infty$. Therefore a negative cosmological
constant will not change the sign of $\eta_N$ since
$B_1(\lambda_k)<0$ for $\lambda_k\le0$.

For $\lambda_k>0$, the $\Phi$'s in \rf{4.55a} are evaluated at
negative arguments $w\equiv -2\lambda_k$. From \rf{4.30} it is
clear that $\Phi^p_n(w)$ blows up for $w\to -1$.
(The function $z+R^{(0)}(z)$ assumes its minimum value 1 at
$z=0$ and increases monotonically for $z>0$.)
This signals that our approximation breaks down for
$\lambda_k\approx 1/2$ or $\bar\lambda_k\approx k^2/2$.
For moderately large values of $\lambda_k$, $B_1(\lambda_k)$
is still negative. As $\lambda_k$ approaches $1/2$ from below,
only the first two terms on the RHS of \rf{4.55a} are important.
It might be that
$B_1$ turns negative then, but this would be in a regime where
our truncation is not reliable any more, and the
sign would even depend on $R^{(0)}$ in general.

At this point a general remark concerning the domain of validity
of our truncation might be in order. In section 3 we showed
that truncations of the form \rf{3.11} with $\wh \Gamma_k=0$
are consistent with the modified Ward identities provided
$Y_k$ is small. For the Einstein--Hilbert truncation we can
evaluate the traces in \rf{3.9} and we can express $Y_k$ in
terms of the functions $\Phi^p_n(w)$. It is clear, therefore,
that $Y_k$ becomes large for $w\to-1$, and that our truncation
cannot account for this regime.

The running of the (dimensionful) cosmological constant itself
is governed by the equation
\be\label{4.64}
\partial_t\bar\lambda_k=
\eta_N \bar\lambda_k
+\frac{1}{2\pi}k^4G_k
\left[10\Phi^1_2(-2\bar\lambda_k/k^2)
     -8\Phi^1_2(0)-5\eta_N\Phi^1_2(-2\bar\lambda_k/k^2)
\right]
\ee
If we switch off the renormalization group improvement for a moment
and set $\eta_N=0$, $\bar\lambda_k=0$ on the RHS of eq.\rf{4.64},
it has the solution
\be\label{4.65}
\bar\lambda_k=\frac{1}{4\pi}\Phi^1_2(0)\bar G(k^4-\Lambda^4)
+\bar\lambda_\Lambda
\ee
We observe the canonical scale dependence $\bar\lambda_k\sim k^4$
which one expects in any naive one--loop calculation: if $\bar\lambda_k$
starts off positive at $k=\Lambda$, its absolute value decreases
when $k$ is lowered until it reaches zero and then $\bar\lambda_k$
becomes negative (for $\Lambda$ large enough). It is obvious that any
attempt to fine--tune $\bar\lambda_\Lambda$ in such a way that
$\dis\lim_{k\to 0}\bar\lambda_k=0$ cannot have a universal meaning
because $\Phi^1_2(0)$ depends on the form of the cutoff.
The evolution equation \rf{4.64} improves on the one--loop
result in two respects: it includes the effect of the running
$G_k$, and via the ``threshold function'' $\Phi^1_2$ it describes
the backreaction of the changing $\bar\lambda_k$ on its
$\beta$--function. In particular, for $\bar\lambda_k<0$ and
$k^2\ll\vert\bar\lambda_k\vert$ the relevant IR cutoff in the graviton
propagator is $\vert\bar\lambda_k\vert$ rather than $k^2$. Then the
graviton modes do not contribute to the running of $\bar\lambda_k$
any longer, and their decoupling is described by the function
$\Phi^1_2(w)$. If, on the other hand, the evolution starts with
$\bar\lambda_k>0$, the threshold functions make the coefficient of the
$k^4$--term in \rf{4.64} even larger, and the running towards zero
is faster than in \rf{4.65}. This effect is counteracted by the
term $\eta_N\bar\lambda_k$
which is negative for $\eta_N<0$. It cannot prevent $\bar\lambda_k$
from overshooting zero, however.

\mysection{Conclusion}
In this paper we proposed a general framework for the treatment
of quantum gravity along the lines of the Wilsonian renormalization
group. We introduced a scale--dependent effective action and we
derived an exact renormalization group equation which describes its
dependence on the built--in infrared cutoff. The effective action
is invariant under general coordinate transformations;
no symmetry violating terms are generated during the evolution.
It satisfies
a set of modified gravitational Ward identities which ensure that,
in the limit of a vanishing cutoff, the conventional Ward identities
are recovered. By virtue of the diffeomorphism--invariance of the
effective action, fairly simple invariant truncations of the space of actions
are sufficient to describe the essential physics in a nonperturbative
way.
The modified Ward identities provide a check for the quality of the
truncations. The evolution equation can be used both for the quantization
of fundamental theories ($\Lambda\to\infty$) and for the evolution
of effective theories ($\Lambda$ finite). It is defined in terms of
manifestly finite, ultraviolet convergent functional traces. The
evolution equation by itself is meaningful even if the action is not
positive definite. In this case
the original euclidean functional integral formulation might be
problematic, and the precise relation between the two approaches
is not entirely clear yet.

As a first application,
we have tested our method within a simple truncation which retains
only the invariants $\int\sqrt{g}R$ and $\int\sqrt{g}$.
Nevertheless, the resulting evolution equations for Newton's
constant and the cosmological constant contain nonperturbative
information. In $2+\varepsilon$ dimensions we found corrections
to the $\beta$--function for $G_k$ and we determined its dependence
on the cosmological constant. In 4 dimensions we saw that the
$\beta$--function for $G_k$ depends on $k$ quadratically, and that
Newton's constant increases at large distances. Within its restricted
domain of validity, this result confirms earlier speculations
by Polyakov \cite{polya} on a possible gravitational antiscreening.

It would be interesting to allow for a more general truncation and
to include  more complicated invariants in the ansatz for $\Gamma_k$.
Not only higher powers of the curvature should be kept but also, and
perhaps more importantly, nonlocal terms must be included
(similar to the 2D induced gravity action $\int R D^{-2}R$,
for instance).
This would lead to a better understanding of quantum gravity in the
extreme infrared, and might help to clarify certain issues in
quantum cosmology. For instance, it has been proposed that quantum
gravitational effects at large distances should be important both
in the context of the dark matter problem \cite{dark} and the
cosmological constant problem \cite{mott,polya}. In fact, it is
quite clear that the nature of the IR divergences, and hence
of the renormalization group flow for $k\to 0$, is quite different
depending on whether $\lambda$ is zero or not \cite{tsam}. In a
perturbative expansion, one of the traces on the RHS of the evolution
equation  consists of graviton loops attached to external graviton
lines. The most singular (for $k\to 0$) diagrams are those which
involve the vertices obtained by expanding $\lambda\int\sqrt{g}$,
because they do not contain any momentum factors. Hence for
$\lambda\ne0$ the renormalization effects should be much stronger
than for $\lambda=0$, and this could eventually drive the
cosmological constant to zero. We hope to come back to this point
elsewhere.

\noindent
Acknowledgement: I would like to thank C.Wetterich for many helpful
discussions.

\end{document}